\documentclass[preprintnumbers]{revtex4}

\usepackage{epsfig}

\begin{document}
\title{Scaling of Yukawa Couplings and Quark Flavor Mixings in the UED Model}

\author{Lu-Xin~Liu\footnote{Talk presented at the Workshop on Discovery Physics at the LHC -Kruger 2010, December 05-10, 2010, and as shall appear in the PoS workshop proceedings.}}
\email[Email: ]{luxin.liu9@gmail.com}
\affiliation{National Institute for Theoretical Physics; School of Physics, University of the Witwatersrand, Wits 2050, South Africa}
\author{A.~S.~Cornell}
\email[Email: ]{alan.cornell@wits.ac.za}
\affiliation{National Institute for Theoretical Physics; School of Physics, University of the Witwatersrand, Wits 2050, South Africa}

\begin{abstract}
The evolution properties of Yukawa couplings and quark mixings are performed for the one-loop renormalization group equations in the Universal Extra Dimension (UED) model.  It is found that the UED model has a substantial effect on the scaling of the fermion masses, including both quark and lepton sectors, whilst the radiative effects on the unitarity triangle is not a sensitive test in this model. Also, for this model, the renormalization invariants $R_{13}$ and $R_{23}$ describe the correlation between the mixing angles and mass ratios to a good approximation, with a variation of the order of $\lambda^4$ and $\lambda^5$ under energy scaling respectively.
\end{abstract}

\preprint{WITS-CTP-67}
\maketitle

\par In particle physics one of the major issues is to explain the fermion mass hierarchy and their mixings. A clear feature of the fermion mass spectrum gives us \cite{Falcone:2001ep,Kuo:2005jt,Liu:2009vh}
\begin{eqnarray}
{m_u} \ll {m_c} \ll {m_t} \; , \;\; 
{m_d} \ll {m_s} \ll {m_b}\; , \;\; 
{m_e} \ll {m_\mu } \ll {m_\tau }\; , \label{eqn:1}
\end{eqnarray}
where a completely satisfactory theory of fermion masses and the related problem of mixing angles is certainly lacking at present. However, there has been considerable effort to understand the hierarchies of these mixing angles and fermion masses in terms of the renormalization group equations (RGE) \cite{Liu:2009vh,Cornell:2010sz,Babu:1987im,Dienes:1998vg,Barger:1992pk}.

\par In general we use RGE as a probe to study the momentum dependence of the Yukawa couplings, the gauge couplings, and the Cabibbo-Kobayashi-Maskawa (CKM) matrix elements themselves. In the Standard Model (SM), the one-loop corrections to the gauge couplings are given by
\begin{eqnarray}
16{\pi ^2}\frac{{d{g_i}}}{{dt}} = {b_i}^{SM}{g_i}^3, \label{eqn:2}
\end{eqnarray}
where ${b_i}^{SM} = (\frac{{41}}{10}, - \frac{{19}}{6}, - 7)$, $t = \ln (\mu /{M_Z})$, and ${M_Z}$ is the $Z$ boson mass. These equations lead directly to the well known gauge unification around ${10^{14}}$ GeV scale.

\par With the commencement of the Large Hadron Collider (LHC), physicists have begun to explore the realm of new physics around the TeV scale. Among these new scenarios the Universal Extra Dimension (UED) model features a tower of Kaluza-Klein (KK) states for each of the SM fields, all of which have access to the extended spacetime manifold \cite{Cornell:2010sz}. In the simplest case, there is a single flat extra dimension of size $R$, compactified on an ${S_1}/{Z_2}$ orbifold. The zero modes of the KK expansion of the SM fields in five dimensional spacetime are identified with the 4-dimensional SM fields.

\par
\begin{figure}[tb]
\begin{center}
\epsfig{file=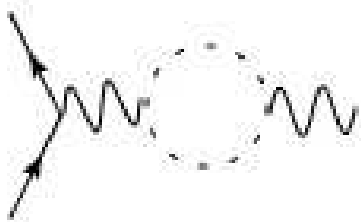,width=.3\textwidth}
\caption{\sl The self-energy interactions of the gauge field with the $A_5^n$ scalar fields.}
\label{fig:2}
\end{center}
\end{figure}

 At each excited KK level the one-loop corrections to the gauge couplings arise from the diagrams exactly mirroring those of the SM ground states \cite{Cheng:1973nv}(however, for the closed fermion one-loop diagrams, one needs to count the contributions from both the left-handed and right-handed KK modes of each chiral fermion to the self-energy of the gauge field), plus new contributions to the self-energy of the gauge boson from the fifth component of the 5D gauge field ${A_M}$ ($M = 0,1,2,3,5$) at each KK excited level as depicted in Fig.\ref{fig:2}

\par Between the scale ${R^{ - 1}}$ where the first KK states are excited and the cutoff scale $\Lambda$, there are finite quantum corrections to the Yukawa and gauge couplings from the $\Lambda R$ number of KK states. Up to the scale ${R^{ - 1}}$ the first step KK excitation occurs, the RG evolution is logarithmic, controlled by the SM beta functions. With increasing energy, that is, as each KK threshold is crossed, new excitations come into play and govern new sets of beta functions until the next threshold is reached. Therefore, to a good approximation, the structure of the one-loop RGE for the gauge couplings are governed by
\begin{eqnarray}
16{\pi ^2}\frac{{d{g_i}}}{{dt}} = [{b_i}^{SM} + (S(t) - 1){{\tilde b}_i}]{g_i}^3, \label{eqn:3}
\end{eqnarray}
where $S(t) = {e^t}{M_Z}R$, and ${{\tilde b}_i} = (\frac{{81}}{{10}},\frac{7}{6}, - \frac{5}{2})$, corresponding to each of the gauge couplings. We can see that the dependence of the gauge couplings on the energy scale drastically changes the normal one-loop running of the gauge couplings, and lowers the unification scale considerably.
\begin{figure}[tb]
\begin{center}
\epsfig{file=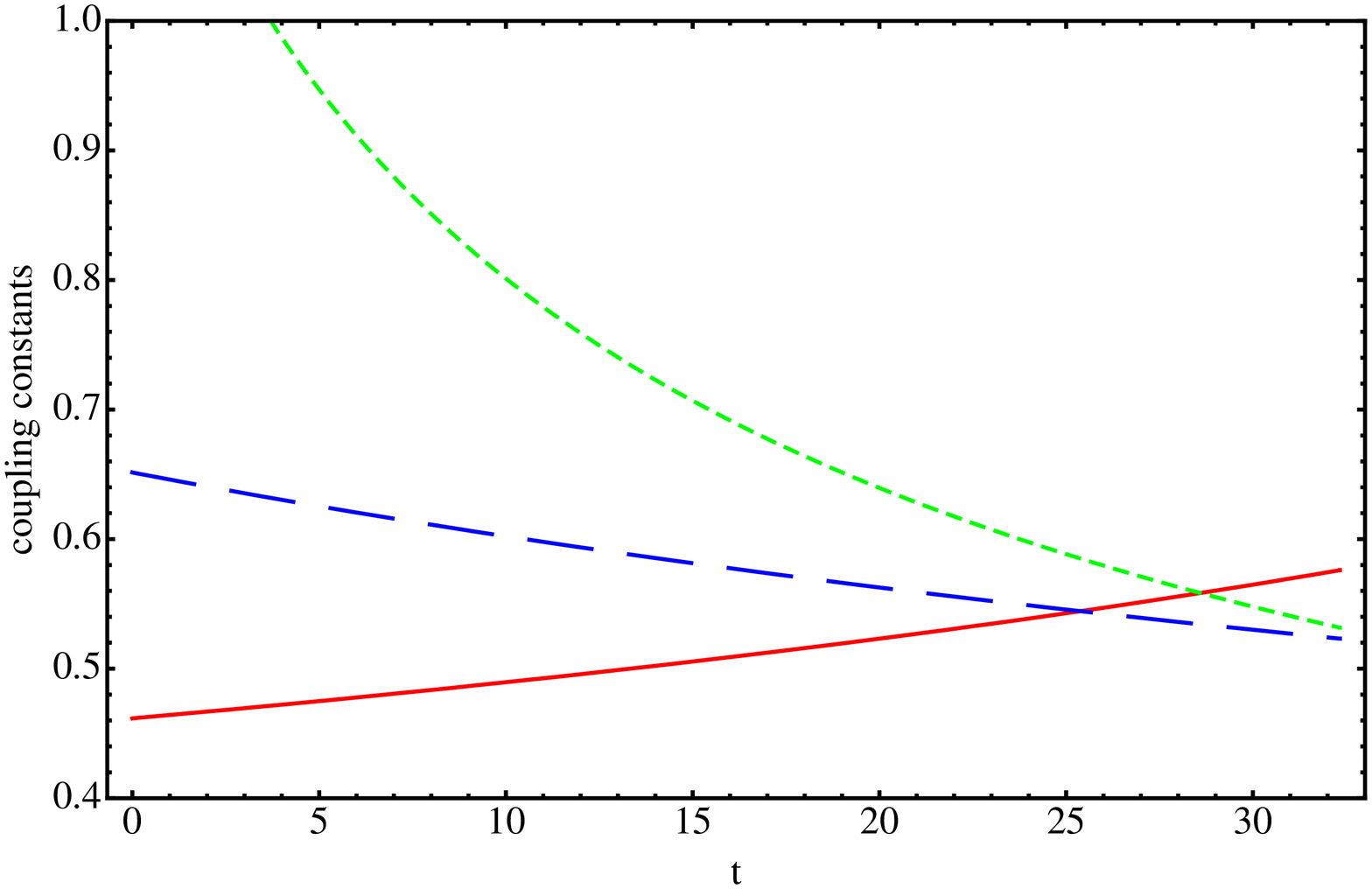,width=.4\textwidth}
\epsfig{file=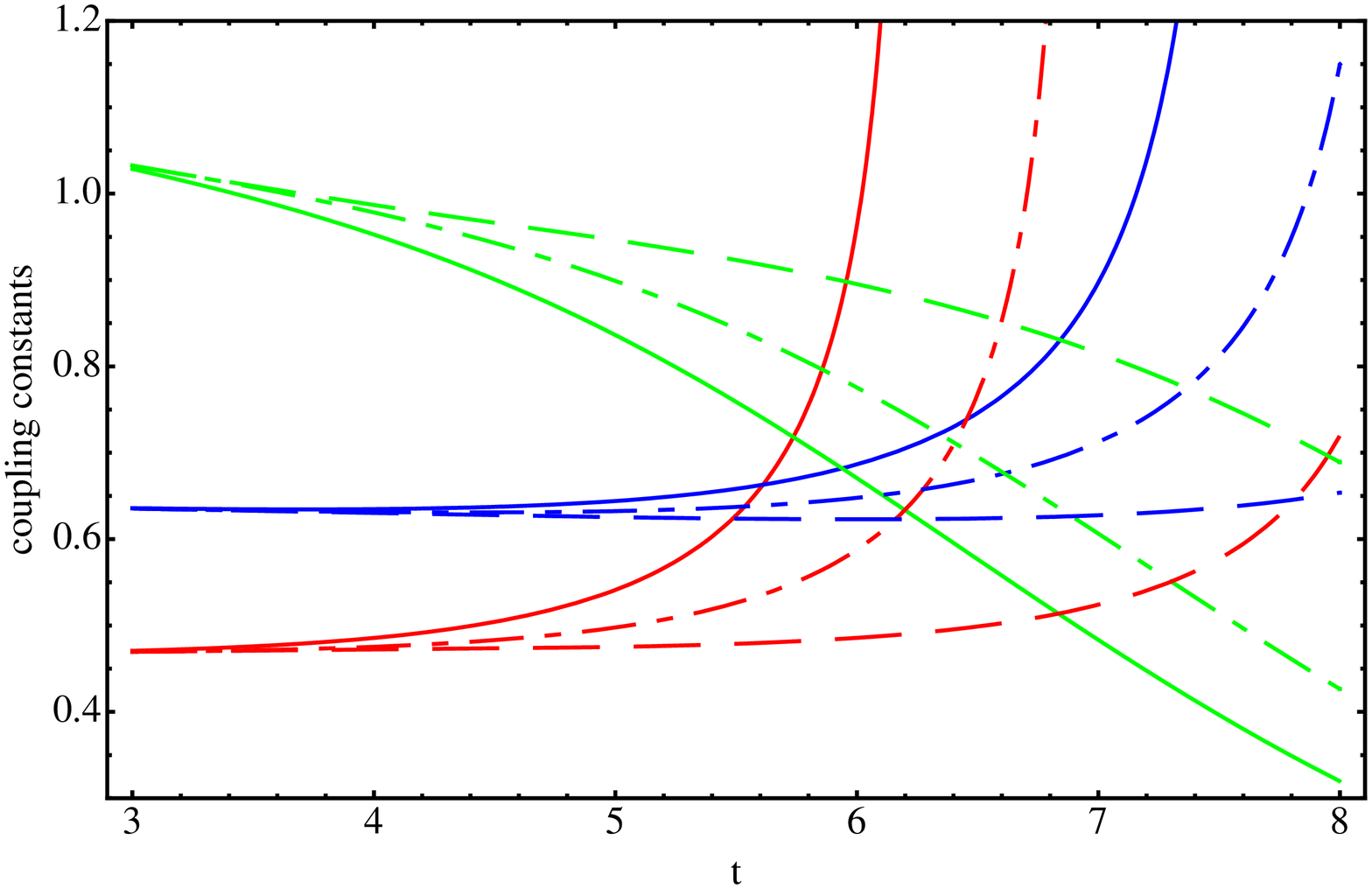,width=.4\textwidth}
\caption{\sl The evolution of the gauge coupling constants in the (left panel) SM, where the dotted line is for $g_3$, the dashed line for $g_2$ and the solid line for $g_1$; and the (right panel) UED model where the solid lines are for the $R^{-1} = 1$ TeV case, the dotted-dashed line is the 2 TeV case and the dashed line is the 10 TeV case. The green lines are for $g_3$, blue for $g_2$ and $g_1$ is red.}
\label{fig:3}
\end{center}
\end{figure}

\par As illustrated in the right panel of Fig.\ref{fig:3}, the extra spacetime dimensions naturally lead to the appearance of grand unified theories at scales substantially below the usual GUT scale. Specifically, for the compactification radii ${R^{ - 1}} = 1,\; 2,\; 10$ TeV, we find the gauge couplings meet at around $\Lambda \sim 30,\; 60,\; 330$ TeV respectively.

\par Given that extra spacetime dimensions induce power law corrections for the gauge couplings, it is natural to ask whether the fermion mass hierarchy might also be explained in our scenario.  In fact, the Yukawa couplings also receive finite one-loop corrections at each KK level and whose magnitudes depend upon the cut off energy scale. Explicitly, the one-loop RGE for the Yukawa couplings take the following form:
\begin{eqnarray}
16{\pi ^2}\frac{{d{Y_U}}}{{dt}} = \beta _U^{SM} + \beta _U^{UED}\; , \;\;
16{\pi ^2}\frac{{d{Y_D}}}{{dt}} = \beta _D^{SM} + \beta _D^{UED}\; , \;\;
16{\pi ^2}\frac{{d{Y_E}}}{{dt}} = \beta _E^{SM} + \beta _E^{UED}\; , \label{eqn:4}
\end{eqnarray}
where the Yukawa coupling beta functions $\beta _U^{SM}$, $\beta _U^{UED}$, $\beta _D^{SM}$, $\beta _D^{UED}$, $\beta _E^{SM}$, and $\beta _E^{UED}$ are defined in \cite{Cornell:2010sz}. The square of the up-type and down-type Yukawa coupling matrices can be diagonalized by using two unitary matrices $U$ and $V$
\begin{eqnarray}
UY_U^\dag {Y_U}{U^\dag } = diag(f_u^2,f_c^2,f_t^2) \; , \;\;
VY_D^\dag {Y_D}{V^\dag } = diag(h_d^2,h_s^2,h_b^2) \; . \label{eqn:5}
\end{eqnarray}
The evolution equations of the eigenvalues $f_u^2,f_c^2,f_t^2$ and $h_d^2,h_s^2,h_b^2$ have the following form:
\begin{eqnarray}
16{\pi ^2}\frac{{d{f_i}^2}}{{dt}} = {f_i}^2[2(2S - 1)T - 2{G_U} + 3S{f_i}^2 - 3S\sum\limits_j  {{h_j }^2} {\left| {{V_{ij }}} \right|^2}]\; , \nonumber\\
16{\pi ^2}\frac{{d{h_j }^2}}{{dt}} = {h_j }^2[2(2S - 1)T - 2{G_D} + 3S{h_j }^2 - 3S\sum\limits_i {{f_i}^2} {\left| {{V_{ij }}} \right|^2}]\; , \label{eqn:6}
\end{eqnarray}
where $i=u,c,t$ and $j=d,s,b$. The Yukawa coupling of the lepton sector can be further re-written as
\begin{eqnarray}
16{\pi ^2}\frac{{d{y_a}^2}}{{dt}} = {y_a}^2[2(2S - 1)T - 2{G_E} + 3S{y_a}^2]\; , \label{eqn:7}
\end{eqnarray}
with ${Y_E} = diag({y_e},{y_\mu },{y_\tau })$. Also, note that
${G_U} = 8g_3^2 + \frac{9}{4}g_2^2 + \frac{{17}}{{20}}g_1^2 + (S - 1)(\frac{{28}}{3}g_3^2 + \frac{{15}}{8}g_2^2 + \frac{{101}}{{120}}g_1^2)\; ,$
${G_D} = 8g_3^2 + \frac{9}{4}g_2^2 + \frac{1}{4}g_1^2 + (S - 1)(\frac{{28}}{3}g_3^2 + \frac{{15}}{8}g_2^2 + \frac{{17}}{{120}}g_1^2)\; ,$
${G_E} = (\frac{9}{4}g_2^2 + \frac{9}{4}g_1^2) + (S - 1)(\frac{{15}}{8}g_2^2 + \frac{{99}}{{40}}g_1^2)\; ,$
$\mathrm{and} \; T = Tr[3Y_U^\dag {Y_U} + 3Y_D^\dag {Y_D} + Y_E^\dag {Y_E}]\; .$ Thus, together with Eq.(\ref{eqn:9}), we have obtained the full set of one-loop coupled RG equations for the Yukawa coupling, the CKM matrix, and the gauge couplings. From these we can obtain the renormalization group flow of all observables related to up and down quark masses and the CKM matrix elements. As such Eqs.(\ref{eqn:3}, \ref{eqn:6}, \ref{eqn:7}, \ref{eqn:9}) constitute a complete set of coupled differential equations for the three families.

\par
\begin{figure}[tb]
\begin{center}
\epsfig{file=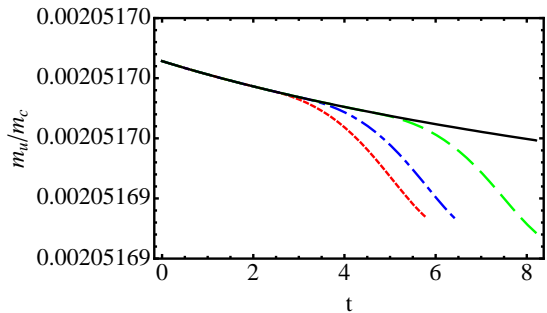,width=.4\textwidth}
\caption{\sl The evolution of the mass ratio $m_u/m_c$, where the solid line is for the SM, the dashed (green) line is the $R^{-1} = 10$ TeV UED case, the dotted-dashed (blue) line is the 2 TeV UED case and the dotted (red) line is the 1 TeV UED case. Note that we shall use the same notation for the figures that follow.}
\label{fig:4}
\end{center}
\end{figure}

 For the hierarchy between the first two light generations, in the leading order approximation, one finds that the running behaviors of the mass ratios are governed by the combination of the third family Yukawa couplings and the CKM matrix elements. This implies that the mass ratios of the first two light generations have a slowed evolution well before the unification scale. Beyond that point new physics would come into play (for example, see Fig.\ref{fig:4}). Quantitatively, similar to the conclusions found in the SM \cite{Barger:1992pk, Fritzsch:1999ee}, here we find the scaling dependence of ${m_d}/{m_s}$ and ${m_e}/{m_\mu }$ is also very slow.
\par
\begin{figure}[tb]
\begin{center}
\epsfig{file=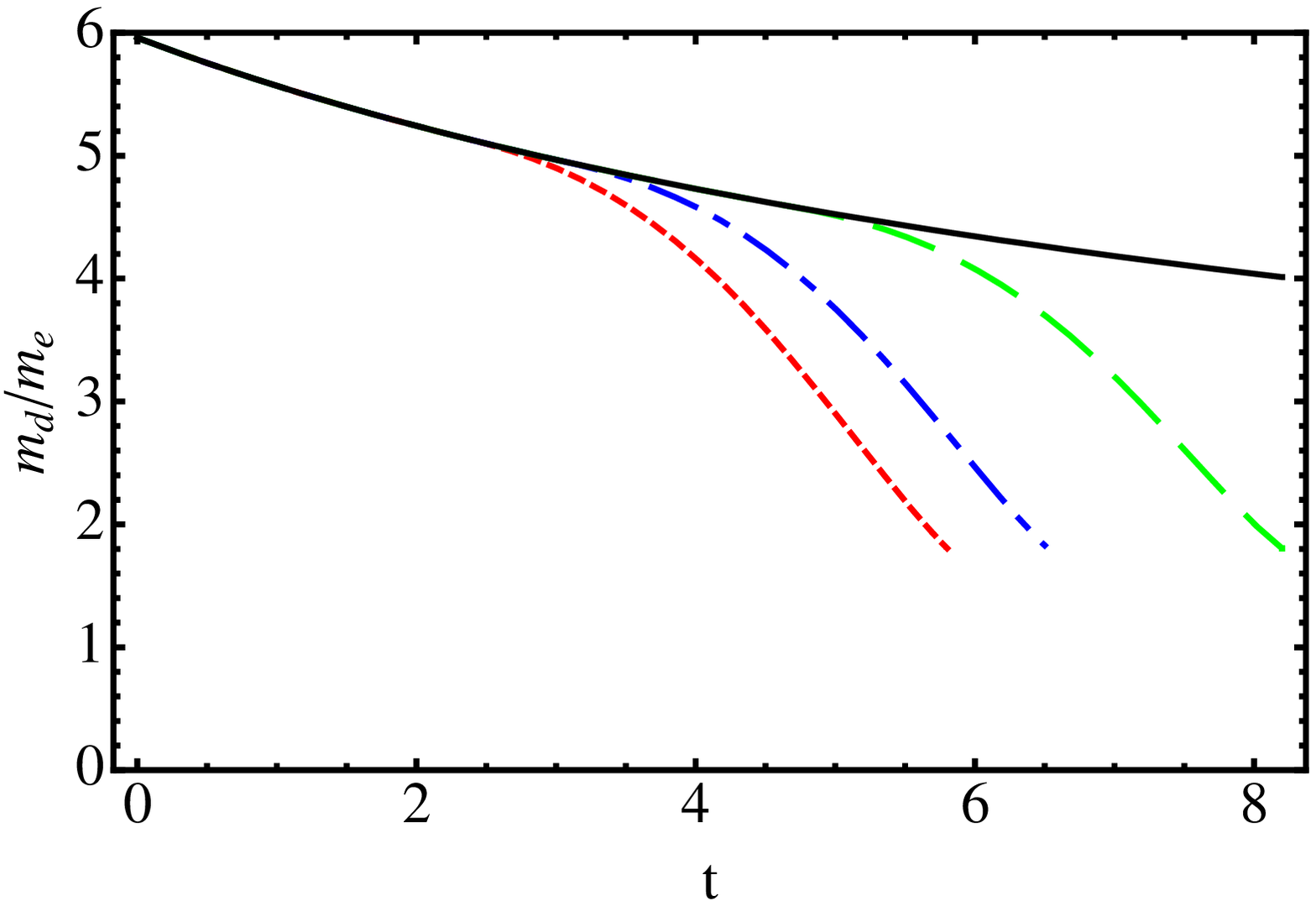,width=.3\textwidth}
\epsfig{file=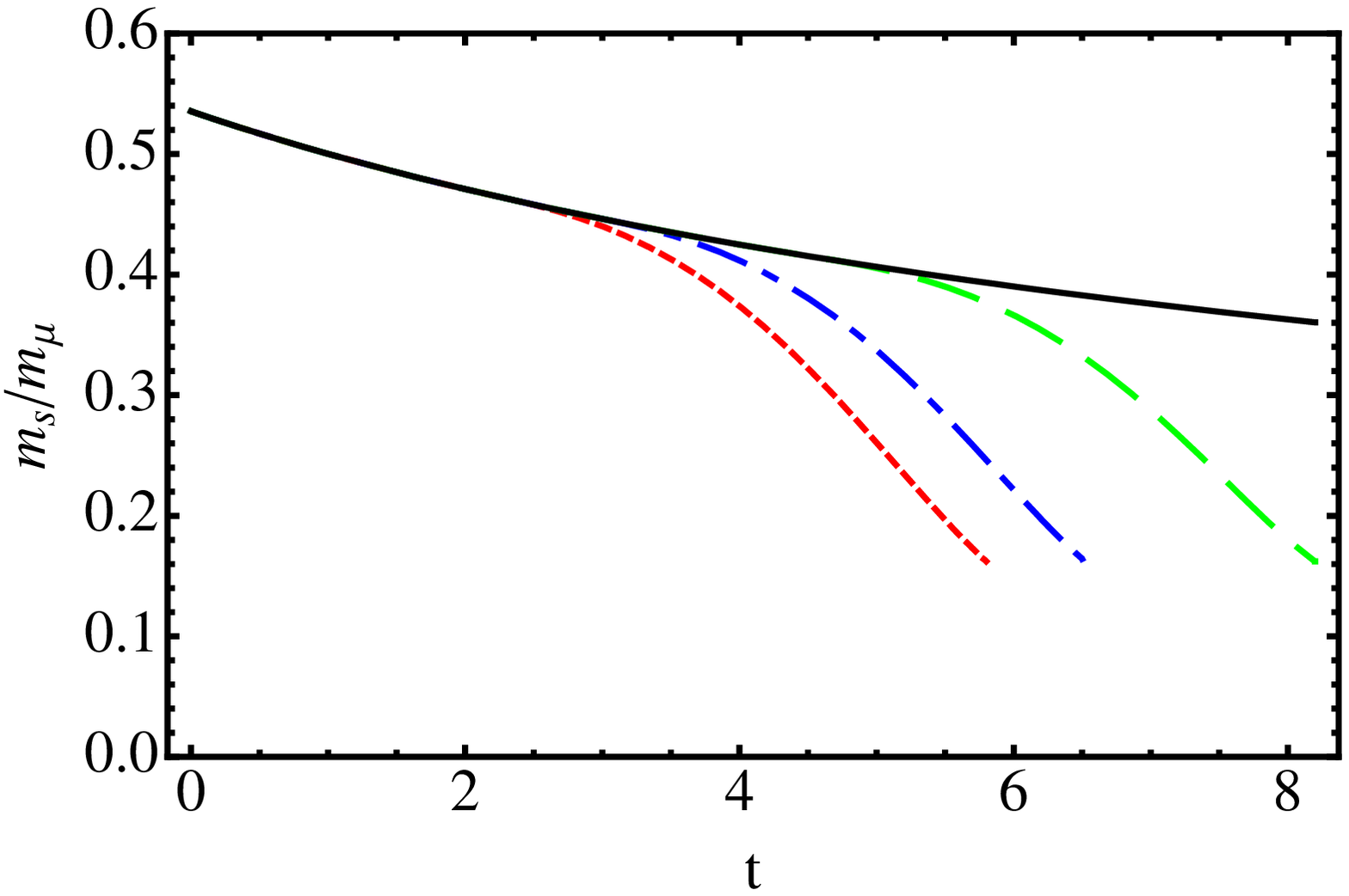,width=.3\textwidth}
\epsfig{file=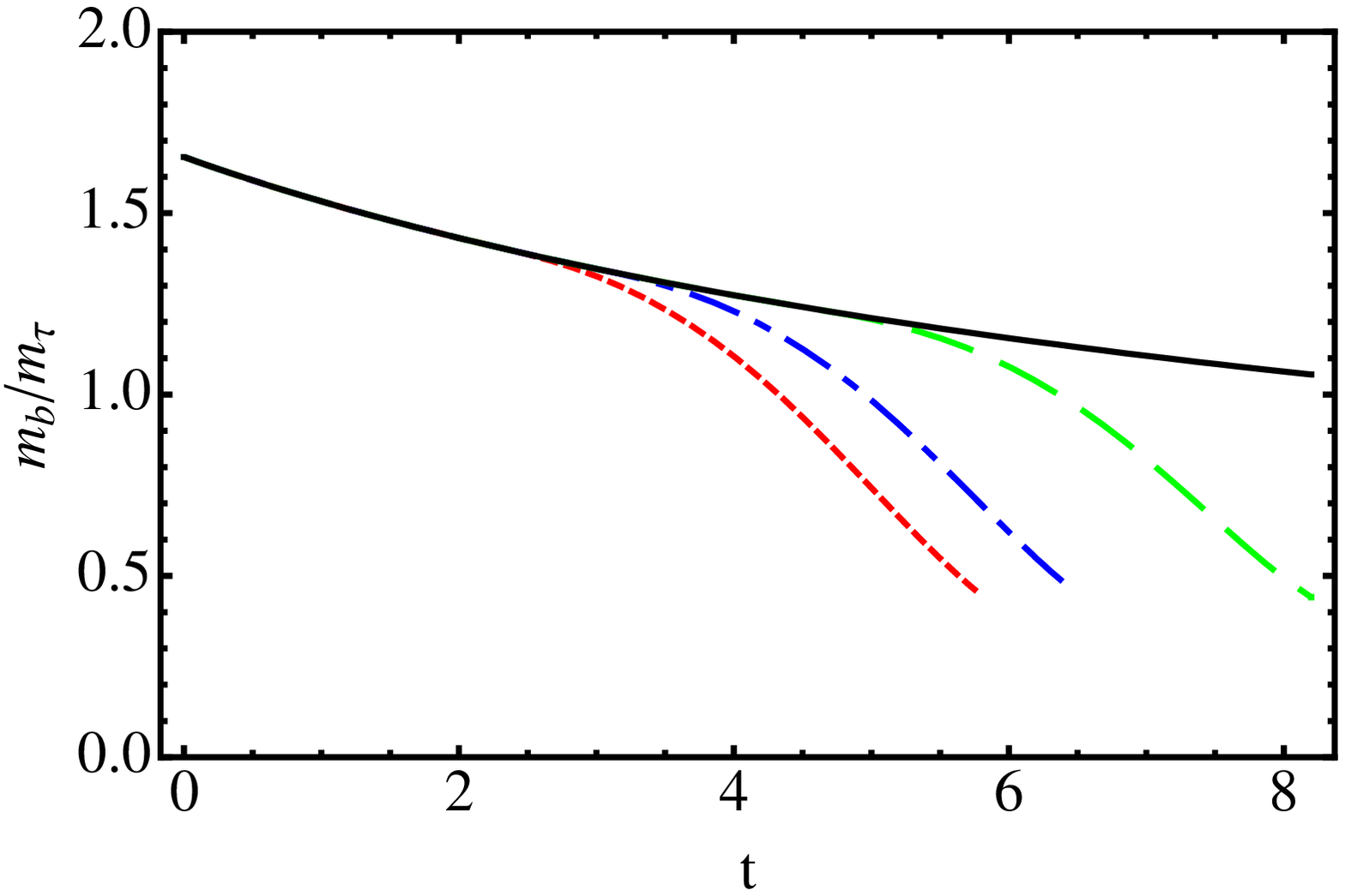,width=.3\textwidth}
\caption{\sl The evolution of the mass ratio (left panel) $m_d/m_e$, (centre panel) $m_s/m_\mu$, and (right panel) $m_b/m_\tau$.}
\label{fig:5}
\end{center}
\end{figure}

 On the other hand, in Grand Unification Theories, such as the $SU(5)$ theory, we place the quark and lepton fields on the same footing when we fill out the field multiplet for the group representation. From the mass matrix relation we have ${m_d} = {m_e}$, ${m_s} = {m_\mu }$, and ${m_b} = {m_\tau }$ at the unification scale. These relations hold such that the differences of their mass values at the electro-weak scale are understood as a running effect.  In the UED model, due to the power law running of the Yukawa couplings, the renormalization effect on these relations can be large. In Figs.\ref{fig:5} we present the numerical analysis of the one-loop calculation of the mass ratios ${m_d}/{m_e}$, ${m_s}/{m_\mu }$, and ${m_b}/{m_\tau }$ respectively.
\par As illustrated, the mass ratios evolve in the usual logarithmic fashion when the energy is below 1TeV, 2TeV, and 10 TeV for the three different cases. However, once the first KK threshold is reached, the contributions from the KK states become more and more significant, at which point their running deviates from their normal orbits and begin to evolve rapidly. As observed, the mass ratios all decrease with increasing energy, which agrees with what is observed in the SM, however, the mass ratios decrease at a much a faster rate. In the UED model the mass ratios for the three families have a sizable variation, which is more than 60\% across the whole range, and this is almost twice as great as that of the SM. This is an interesting feature that distinguishes these two models. Also, a flavor dependent interaction can be introduced to the Yukawa couplings and this flavor dependence need not be very strong, since the power law effects of RGE can easily amplify the effects of even a relatively mild flavor dependence \cite{Dienes:1998vg}. Therefore, due to the fast power law running the unification of the Yukawa couplings is very desirable, where this feature has the potential to address the problem of fermion mass hierarchy.

\par From Eq.(\ref{eqn:5}) it follows that the CKM matrix describing the quark flavor mixing in the charged current is given by ${V_{CKM}} = U{V^\dag }$, where from Eqs.(\ref{eqn:4}, \ref{eqn:6}) the RGE of the CKM matrix elements are:
\begin{eqnarray}
16{\pi ^2}\frac{{d{{\left| {{V_{ij}}} \right|}^2}}}{{dt}} &=& S(t)\left\{3{\left| {{V_{ij}}} \right|^2}({f_i}^2 + {h_j}^2 - \sum\limits_k {{f_k}^2} {\left| {{V_{kj}}} \right|^2} - \sum\limits_k {{h_k}^2} {\left| {{V_{ik}}} \right|^2}) - 3{f_i}^2\sum\limits_{k \ne i} {\frac{1}{{{f_i}^2 - {f_k}^2}}} \Big(2{h_j}^2{\left| {{V_{kj}}} \right|^2}{\left| {{V_{ij}}} \right|^2} \right.\nonumber \\
&&\hspace{1cm}\left.+ \sum\limits_{l \ne j}{h_l}^2{V_{iklj}}\Big) - 3{h_j}^2\sum\limits_{l \ne j} {\frac{1}{{{h_j}^2 - {h_l}^2}}} (2{f_i}^2{\left| {{V_{il}}} \right|^2}{\left| {{V_{ij}}} \right|^2} + \sum\limits_{k \ne i}{f_l}^2{V_{iklj}})\right\}, \label{eqn:9}
\end{eqnarray}
where ${V_{iklj}} = 1 - {\left| {{V_{il}}} \right|^2} - {\left| {{V_{kl}}} \right|^2} - {\left| {{V_{kj}}} \right|^2} - {\left| {{V_{ij}}} \right|^2} + {\left| {{V_{il}}} \right|^2}{\left| {{V_{kj}}} \right|^2} + {\left| {{V_{kl}}} \right|^2}{\left| {{V_{ij}}} \right|^2}$.

\par
\begin{figure}[tb]
\begin{center}
\epsfig{file=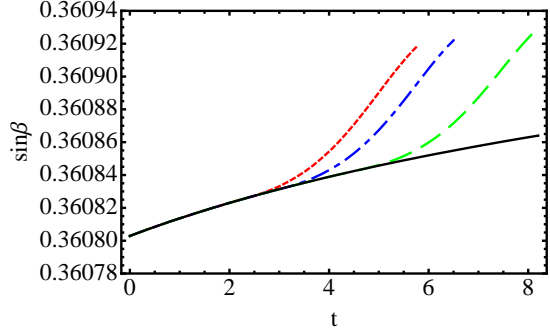,width=.4\textwidth}
\caption{\sl The evolution of $\sin\beta$.}
\label{fig:8}
\end{center}
\end{figure}

 Since the mixing matrix ${V_{CKM}}$ satisfies the full unitary conditions, we have the following constraint
\begin{eqnarray}
{V_{ud}}V_{ub}^* + {V_{cd}}V_{cb}^* + {V_{td}}V_{tb}^* = 0 \; , \label{eqn:10}
\end{eqnarray}
that is, we have a triangle in the complex plane, composed of three sides with lengths $\left| {{V_{ud}}} \right|\left| {V_{ub}^*} \right|$, $\left| {{V_{td}}} \right|\left| {V_{tb}^*} \right|$, and $\left| {{V_{cd}}} \right|\left| {V_{cb}^*} \right|$. The area $A$ is related to the Jarlskog rephasing invariant parameter $J$ through the relation $J = 2 A$. Therefore, we can identify its three inner angles $\alpha$, $\beta$ and $\gamma$ from the area and its sides
\begin{eqnarray}
\sin \beta  = \frac{J}{{\left| {{V_{td}}} \right|\left| {V_{tb}^*} \right|\left| {{V_{cd}}} \right|\left| {V_{cb}^*} \right|}} \; , \;\;
\sin \gamma  = \frac{J}{{\left| {{V_{ud}}} \right|\left| {V_{ub}^*} \right|\left| {{V_{cd}}} \right|\left| {V_{cb}^*} \right|}}\; , \label{eqn:11}
\end{eqnarray}
and $\alpha  = \pi  - \beta  - \gamma$. The shape of the unitarity triangle can be deemed as an important tool to exploring new symmetries or other interesting properties that give a deeper insight into the physical content of new physics models. So it is of great interest to find models in which the CKM matrix might have a simple, special form at asymptotic energies. In Fig. \ref{fig:8} we plot the evolution of the inner angle from the electroweak scale to the unification scale by using the one-loop RGE for the UED model, and demonstrate that the angle has a small variation against radiative corrections.
\par More specifically, the relative deviation for $\sin \beta$ is only up to 0.01\% in the whole range from ${M_Z}$ to the GUT scale. Similar analysis can also be found for the angles $\alpha$ and $\gamma$. This result makes sense, since both the triangle's sides and area become larger and larger when the energy scale increases, the unitarity triangle (UT) is only rescaled and its shape does not change much during the RG evolution. The fact that inner angles are rather stable against radiative corrections indicates that it is not possible to construct an asymptotic model with some simple, special form of the CKM matrix from this simple scenario. The stability against radiative corrections suggests that the shape of the UT is almost unchanged from RGE effects. In this regards, the UT is not a sensitive test of this model in current and upcoming experiments.

\par On the other hand, in the quark sector, both the mass ratios and mixing parameters exhibit rather large hierarchies. At the electroweak scale the observed pattern of fermion masses and mixings
\begin{eqnarray}
{\theta _{13}} \sim \frac{{{m_d}}}{{{m_b}}} \; , \;\;\; {\theta _{23}} \sim \frac{{{m_s}}}{{{m_b}}} \; , \label{eqn:12}
\end{eqnarray}
does not look accidental. In \cite{Liu:2009vh} a set of renormalization invariants is constructed
\begin{eqnarray}
{R_{13}} = \sin (2{\theta _{13}})\sinh \left[\ln \frac{{{m_b}}}{{{m_d}}}\right] \sim constant \; , \;\;
{R_{23}} = \sin (2{\theta _{23}})\sinh \left[\ln \frac{{{m_b}}}{{{m_s}}}\right] \sim constant \; . \label{eqn:13}
\end{eqnarray}
These invariants exhibit explicitly the correlation between quark flavor mixings and mass ratios from the electroweak scale to the GUT scale in the context of the SM, the double Higgs Model, and the Minimal Supersymmetric Standard Model. The well known empirical relations Eq.(\ref{eqn:12}) at the electroweak scale can thus be understood as the result of renormalization evolution toward the infrared point, where the low energy limits will naturally lead to the correlations of Eq.(\ref{eqn:12}). This suggests that relations between mass ratios and mixing angles are dynamical in origin, where it is found that the scale dependence of these quantities for general three flavor mixing follows these invariants up to the GUT scale closely. We quantitatively analyze these quantities in the UED model from the weak scale to the GUT scale.
\begin{figure}[tb]
\begin{center}
\epsfig{file=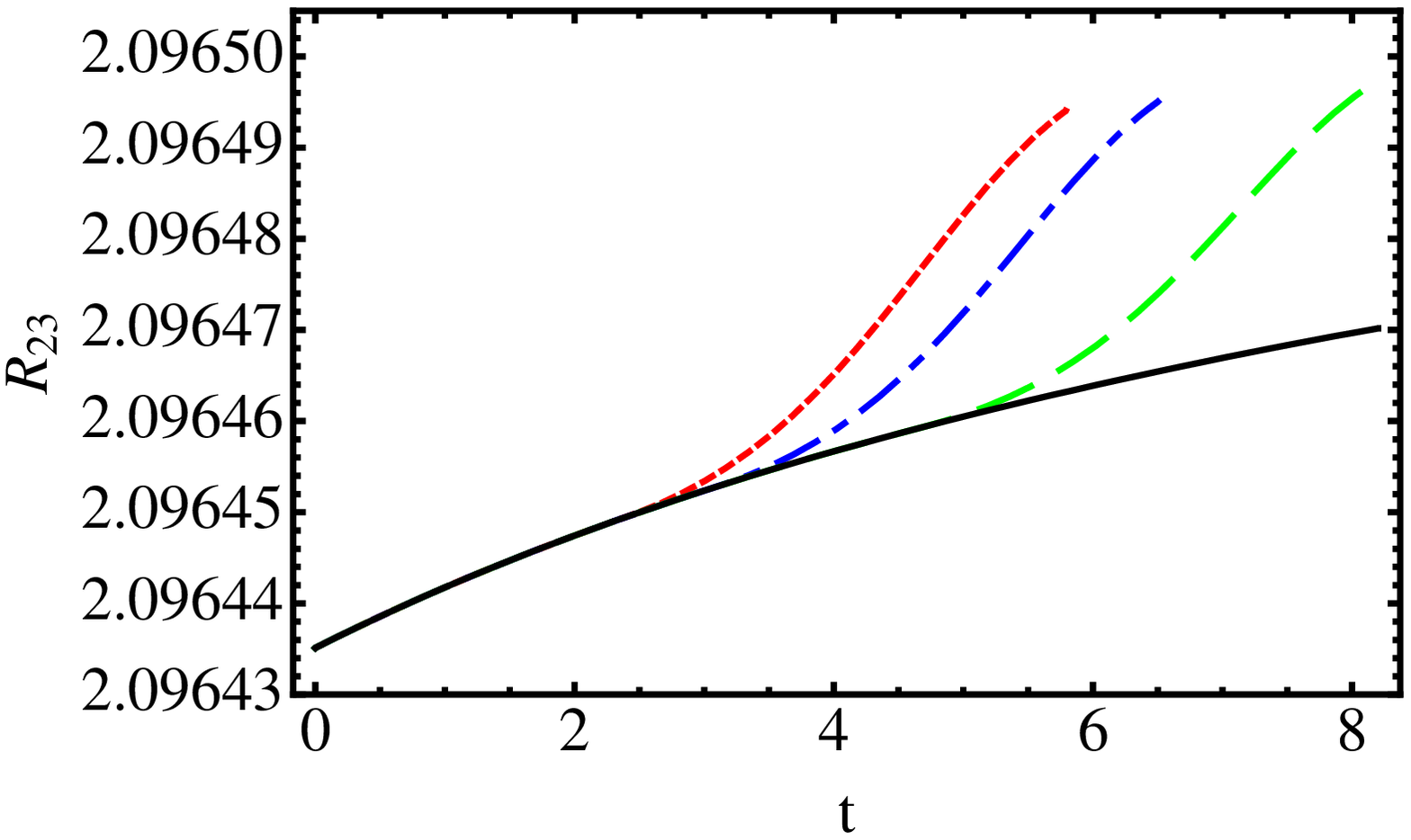,width=.4\textwidth}
\epsfig{file=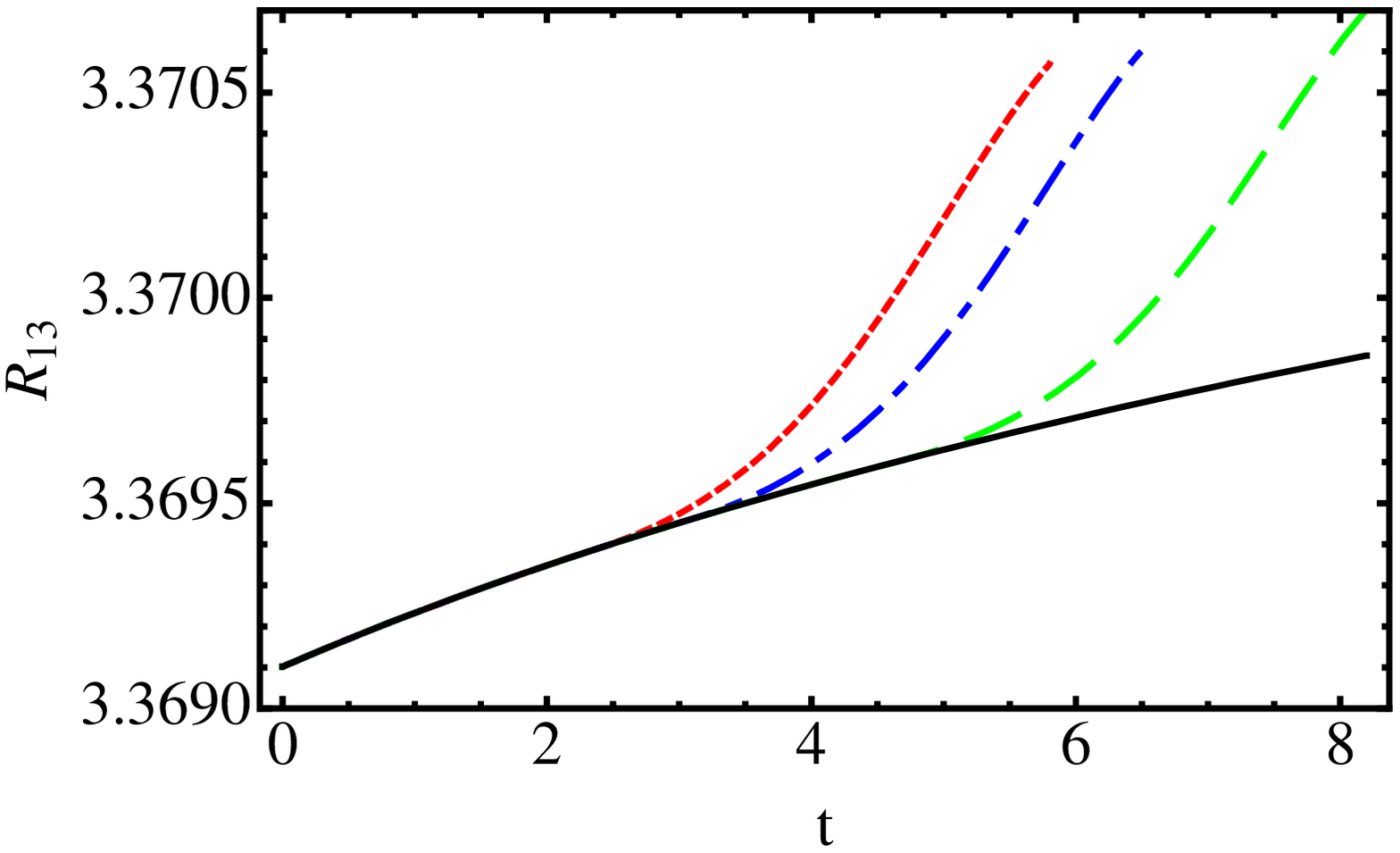,width=.4\textwidth}
\caption{\sl The evolution of the (left panel) $R_{23}$, and (right panel) $R_{13}$.}
\label{fig:9}
\end{center}
\end{figure}
\par As an exemplary example, we take the input values from \cite{ Cornell:2010sz} and plot their evolution. As depicted in Fig.\ref{fig:9}, the energy dependence of ${R_{13}}$ and ${R_{23}}$ are weak. Since the increase of the mixing angles are compensated by the deviation of the mass ratios, the net RGE effect on ${R_{13}}$ and ${R_{23}}$ is not considerable; it only counts up to a relative variation of order ${\lambda ^4}$ and ${\lambda ^5}$ respectively.

\par To conclude, UED models with compactification radius near the TeV scale promises exciting phenomenology for collider physics. It is found that the evolution of the gauge couplings has a rapid variation in the presence of the KK modes and this leads to a much lower unification scale than the SM. As discussed, due to the power law running of the Yukawa couplings, the rapid decrease of the Yukawa couplings with energy is in clear contrast to the slow logarithmic running predicted by the SM. The UED model has substantial effects on the hierarchy between the quark and lepton sectors and provides a very desirable scenario for grand unification. However, we also find that the radiative effect on the UT is not great and not a sensitive test of this model, as the UT is rather stable under the evolution. Finally, to a good approximation, the renormalization invariants ${R_{13}}$ and ${R_{23}}$ describe the correlation between the mixing angles and mass ratios very well, with a variation no more than the order ${\lambda ^4}$ under energy scaling.


\begin{thebibliography}{99}
\bibitem{Falcone:2001ep}
  D.~Falcone,
  Int.\ J.\ Mod.\ Phys.\  A {\bf 17}, 3981 (2002);

\bibitem{Kuo:2005jt}
  T.~K.~Kuo and L.~X.~Liu,
  arXiv:hep-ph/0511037.

\bibitem{Liu:2009vh}
  L.~X.~Liu,
  Int.\ J.\ Mod.\ Phys.\  A {\bf 25}, 4975 (2010)
  [arXiv:0910.1326 [hep-ph]].

\bibitem{Cornell:2010sz}
  A.~S.~Cornell and L.~X.~Liu,
  Phys.\ Rev.\  D {\bf 83}, 033005 (2011)
  [arXiv:1010.5522 [hep-ph]].

\bibitem{Babu:1987im}
  K.~S.~Babu,
  Z.\ Phys.\  C {\bf 35}, 69 (1987).

\bibitem{Dienes:1998vg}
  K.~R.~Dienes, E.~Dudas and T.~Gherghetta,
  Nucl.\ Phys.\  B {\bf 537}, 47 (1999).

\bibitem{Barger:1992pk}
  V.~D.~Barger, M.~S.~Berger and P.~Ohmann,
  Phys.\ Rev.\  D {\bf 47}, 2038 (1993).

\bibitem{Cheng:1973nv}
  T.~P.~Cheng, E.~Eichten and L.~F.~Li,
  Phys.\ Rev.\  D {\bf 9}, 2259 (1974).

\bibitem{Fritzsch:1999ee}
  H.~Fritzsch and Z.~z.~Xing,
  Prog.\ Part.\ Nucl.\ Phys.\  {\bf 45}, 1 (2000).
\end{thebibliography}
\end{document}